\documentclass[onecolumn,pdflatex,nofootinbib,superscriptaddress,floatfix]{revtex4-1}
\usepackage{graphicx}
\usepackage{bm}
\usepackage{placeins}
\usepackage{xspace}
\usepackage{bm,amsmath,amssymb,amsfonts,gensymb}
\usepackage{comment}
\usepackage{hyperref}
\hypersetup{
    colorlinks=true,
    linkcolor=blue,
    citecolor=blue,
    filecolor=blue,      
    urlcolor=blue,
}
\usepackage{cleveref}

\usepackage[russian,english]{babel}

\newcommand{\be}{\begin{equation}}
\newcommand{\ee}{\end{equation}}

\newcommand{\eq}[1]{\begin{align} #1 \end{align}}

\usepackage{color}

\usepackage[english]{babel}
\begin{document}

  \begin{flushleft}
 \hfill \parbox[c]{40mm}{CERN-TH-2022-171}
\end{flushleft}

\title{
Circular orbits of  test particles interacting with massless linear scalar field of the naked singularity  
 }
\author{O.~S.~Stashko}
\affiliation{Frankfurt Institute for Advanced Studies, Frankfurt am Main, Germany}
\affiliation{Goethe Universität, Max-von-Laue Str. 1, Frankfurt am Main, 60438, Germany}
\affiliation{CERN, Theoretical Physics Department, CH-1211 Geneva 23, Switzerland}
\affiliation{Taras Shevchenko National University of Kyiv, Ukraine}

\author{V.~I.~Zhdanov}
\affiliation{Taras Shevchenko National University of Kyiv, Ukraine}
\affiliation{Igor Sikorsky Kyiv Polytechnic Institute, Ukraine}

\date{\today}

\pacs{11}

\begin{abstract}
 We study effects of the particles coupling with scalar field (SF) on the distribution of stable circular orbits (SCO) around the naked singularity  described by the well-known Fisher-Janis-Newman-Winicour  solution (F/JNW).  The power-law and exponential models of the particle--SF interaction  are analyzed. The focus is on the non-connected SCO distributions. A method is used that facilitates numerical studies of the SCO regions in the case of the general static spherically symmetric metric. In case of F/JNW, we show that coupling  between particles and SF   can essentially complicate the topology of the SCO distributions. In particular, it can lead to new non-overlapping SCO regions, which are separated  by unstable orbits and/or by regions where the circular orbits do not exist.

\end{abstract}

\maketitle

\section{Introduction}
\noindent 
Static spherically symmetric configurations of General Relativity (GR) with the linear massless scalar field (SF) are described by the exact analytic solution of Einstein-SF equations, which was firstly derived in 1948 by Fisher \cite{Fisher} and  later rediscovered in the other coordinates by Janis, Newman and Winicour \cite{JNW}. We  will refer  to this solution as F/JNW. Subsequently,  Wyman \cite{Wyman}  obtained this solution within a more general consideration; see also  \cite{Agnese:1985xj,Roberts:1993re} and  comments in \cite{Virbhadra_1997, Abdolrahimi:2009dc}. The F/JNW solution is asymptotically flat, it contains the globally naked singularity (NS) in the center \cite{Virbhadra2etal-1997}, i.e. it can be observed from the infinity.  There are analytic generalizations of F/JNW in case of the  higher dimensions \cite{Xanthopoulos:1989kb,Abdolrahimi:2009dc} and in case of several linear massless SFs \cite{ZS}. 

The picture of the stable circular orbits (SCO) of  test bodies in case of  F/JNW metric has been firstly studied by Chowdhury et al.  \cite{Chowdhury_2012} who revealed the possibility of  non-connected SCO regions, which are separated by unstable circular orbits (UCO). This is an important feature that  distinguishes F/JNW  from the case of  ordinary  Schwarzschild black holes, where there is the region of non-existence of circular orbits (NECO)  near  the center, then the  UCO ring and then the outer (unbounded) region of SCO. Images of  thin accretion disks related to the circular orbits distributions (COD) in case of  F/JNW solutions have been  considered in \cite{Gyulchev_2019,Gyulchev_2020}.  

The unusual  structure of COD  around relativistic objects is not a prerogative of F/JNW; the uncomplete list of papers includes, e.g., 
\cite{Pugliese_2011, Pugliese_2013, Vieira_2014, Meliani_2015, Stuchl_k_2015,  2018Stashko, 2019Stuchlik,
 Dymnikova_2019,Slany_2020,Stashko_Zhdanov_2019a,SZA2}.   As for the configurations with scalar fields, apart from F/JNW, the non-connected SCO distributions have been demonstrated in the case of massive SF  \cite{Stashko_Zhdanov_2019a} and in case of nonlinear SFs with the monomial potential \cite{ZS,SZA2}.  The above cases mainly  deal with NSs, however, the non-connected SCO distributions can arise in presence of the event horizon as well; see, e.g.,  \cite{2018Stashko} dealing with the "Mexican hat" SF potential. 
 
 We note that the investigation of COD  is an important tool to study  strong gravitational fields around compact astrophysical objects, because this is the  first step to figure out the  properties of real accretion discs and their observational properties.  The purpose of such studies is to screen out theories alternative to GR, or, conversely, to look for putative signals of a “new physics” that can be verified in observations. Anyway,  investigations of such exotic structures as NSs with unusual accretion disks is interesting at least because they make the standard black hole paradigm  falsifiable.

The works \cite{Chowdhury_2012,Gyulchev_2019,Gyulchev_2020, Pugliese_2011, Pugliese_2013, Vieira_2014, Meliani_2015, Stuchl_k_2015,  2018Stashko, 2019Stuchlik,
 Dymnikova_2019,Slany_2020,Stashko_Zhdanov_2019a,SZA2} deal with the geodesic particle motion; no additional particle interaction is introduced. In the present paper we focus on the non-connected SCO distributions under assumptions that the massive particles   interact with SF. Analogous interactions naturally arise in the scalar-tensor theories \cite{BransDicke1961,Damour1992,Damour-1993} and in the $f(R)$-gravity after transition to the Einstein frame (see, e.g.,  \cite{DeFelice2010,Shtanov:2021}). There are hard limitations on the particles--SF  interaction imposed by the weak-field gravitational experiments  \cite{Will2014}. Nevertheless,  NS is a  very special object with strong gravitational field and we wonder whether coupling of particles with SF in the  vicinity of NS can lead to  new observational features that can be used
to differentiate between ordinary black hole and NS.

We investigate the effects in the SCO structure due to  various  couplings of the particles with SF   
 within simple toy-models. The particles move in the external  gravitational and scalar fields of the F/JNW naked singularity.  In Section \ref{basic} we write down the basic equations of motion of the particle.

 In Section  \ref{sFJNW} we consider two types of the coupling, which are applied to the equations of the particle motion in the fields of  F/JNW solutions.   We present here the list of possible distributions of the circular orbits; in particular, we present those, which do not occur in case of usual geodesic motion of the particles. 
In Section \ref{discussion} we discuss the results.

\section{Basic relations in case of coupling of particles with SF}\label{basic}
\noindent We consider the motion of the test   particles interacting with SF $\phi $ in the four dimensional space-time of General Relativity endowed with metric $g_{\alpha \beta } $. We assume that the particles have small masses, which makes it possible to neglect the mutual interaction between the particles and their feedback on the gravitational and scalar fields.

For one particle  we assume the action\footnote{Units: $c= 8\pi G=1;$ the metric signature is $(+ - - -)$.} 
\begin{equation} \label{action_particles} S_{m} =-m \int ds \, \psi (\phi (x)),\quad ds=\sqrt{g_{\alpha \beta } (x)\frac{dx^{\alpha } }{d\tau } \frac{dx^{\beta } }{d\tau } } d\tau ;\quad x\equiv x(\tau );           
\end{equation} 
where trajectory $x(\tau )$ is parameterized by a  parameter $\tau $; $\psi(\phi)=1+\xi(\phi)$ is responsible for the interaction of particles with SF. We assume $\psi>0$ in order to avoid negative masses of the particles.  The value  $\xi$ is expected to be sufficiently small in the weak-field systems. However, we look for effects of the interaction in the strong field region, where $\xi$ may be comparable to unity.

The aim of this paper is to show that an additional coupling between particles and SF (besides the indirect interaction through the gravitational field depending on SF by means of the Einstein equations) can essentially complicate the topology of the SCO distributions, in addition to complications due to possible introduction of the non-linear self-interaction SF potentials. This is motivated, in particular, by a number of models, in which  the coupling arises after the transition from Jordan to Einstein frame  either in the Brans-Dicke theory or in the $f(R)$-gravity \cite{BransDicke1961,Damour1992,Damour-1993,DeFelice2010,Shtanov:2021}. In these cases the interaction with SF is simply a result of the conformal transformation of the metric; but this  leads also to changes of the  SF  self-interaction. However, in this paper we are looking for  potentially interesting qualitative coupling effects apart from from those due to the SF potential, which manifest themselves in the strong gravitational field of a naked singularity.
There are various ways to introduce the coupling into the particle action, and we see no physical reason to prefer one model over another. In this situation, the way out is to consider a number of examples with different couplings. Some examples will be considered below in subsections \ref{ssect:Power-law coupling} and \ref{exp-coupling}.  Physically, the coupling may be, e.g., due to some dependence of  masses or interaction constants upon SF.

 The equations of motion for the massive test particle trajectory as a function of the proper time $s$ obtained from  action (\ref{action_particles}) are 
\begin{equation} \label{Eqom} 
\frac{d}{ds} \left(\psi g_{\alpha \beta } \frac{dx^{\beta } }{ds} \right)=\frac{\partial \psi }{\partial x^{\alpha } } +\frac{\psi }{2} \frac{\partial g_{\beta \gamma } }{\partial x^{\alpha } } \frac{dx^{\beta } }{ds} \frac{dx^{\gamma } }{ds} ,        \end{equation} 
  $\psi \equiv \psi [\phi (x(s))]$. Equation (\ref{Eqom}) is equivalent to the geodesic equation for the conformal metric $\tilde g_{\mu\nu}=\psi^2 g_{\mu\nu}$  with the canonical parameter $\tilde s: d\tilde s=\psi ds$.  The trajectories of photons and the other massless particles are the same  under the conformal transformation as for $\psi=1$.

 The equivalent (explicitly covariant) form of this equation of motion is
  \begin{equation} \label{Eqom_1} 
  \frac{\delta}{ds}  \frac{dx^{\beta } }{ds} =\left(  g^{\beta  \mu}-  \frac{dx^{\beta } }{ds} \frac{dx^{\mu } }{ds}\right)\frac{\partial \ln \psi }{\partial x^{\mu } }\,,
  \end{equation} 
were $\delta/ds$ stands for the covariant differentiation (with respect to $g_{\mu\nu}$) along the particle trajectory.  

It is easy to check that in virtue of (\ref{Eqom}) or (\ref{Eqom_1})   
 \begin{equation}
    \frac{d}{ds} \left[\psi ^{2} \left(g_{\mu \nu } \frac{dx}{ds} ^{\mu } \frac{dx}{ds} ^{\nu } -1\right)\right]=0 \,   .
 \end{equation}
This equation is  consistent with usual  normalization 
\begin{equation}\label{normalization1}
g_{\mu \nu } \frac{dx}{ds} ^{\mu } \frac{dx}{ds} ^{\nu } =1 .
\end{equation}
Now we move on to the metric of the static spherically symmetric space-time 
\begin{equation} \label{metric} 
dS^{2} =A(r) dt^{2} -B(r) dr^{2} -[R(r)]^{2} \left(d\theta ^{2} +\sin ^{2} \theta d\varphi ^{2} \right) .
\end{equation} 
Equation (\ref{normalization1}) in the   plane $\theta \equiv \pi /2$ yields
\begin{equation}
 A\left(\frac{dt}{ds} \right)^{2} -B \left(\frac{dr}{ds} \right)^{2} -R^{2} \left(\frac{d\varphi }{ds} \right)^{2} =1  
 \label{normalization} 
\end{equation}
Using (\ref{Eqom}) we have integrals of motion 
\[\psi A\frac{dt}{ds } =p_{0} ,\quad
\psi R^{2} \frac{d\varphi }{ds } =L,\] 
 $p_0$ and $L$ are constants.

After substitution into (\ref{normalization}) we get
\begin{equation} \label{GrindEQ__8_} 
\psi^{2} AB\left(\frac{dr}{ds } \right)^{2} =p_0^2-U_{eff} ,                               \end{equation} 
where the effective potential is 
\[ U_{eff}=A\left(\frac{L^{2} }{R^{2} } +\psi ^{2} \right).\] 
This potential governs properties of COD. Minima of $U_{eff}$ for fixed $L$ correspond to SCO, maxima -- to UCO. The regions where there is no minima or maxima for all $L$ correspond to NECO.
For SCO of radius $r_0$ we have $p_0^2=U_{eff}$ and $U_{eff}$ reaches minimum at $r=r_0$, i.e.  
\begin{equation} \label{extremum_U} U'_{eff}= \left[A'\left(\frac{L^{2} }{R^{2} } +\psi ^{2} \right)-\frac{2L^{2}R' A}{R^{3} } +2A\psi \psi '\right]=0, \quad r=r_0.       \end{equation} 
Consider function
\begin{equation} \label{L-of-r} 
{\rm \tilde{L}}^{2} (r)\equiv -\frac{R^{3} \psi \left(A'\psi +2A\psi '\right)}{D}  ,\quad       D(r)=RA'(r)-2 A R'(r) .      
\end{equation} 
 Then condition \eqref{extremum_U} can be written as $L^{2} ={\rm \tilde{L}}^{2} (r_0)$, $r_0$ being a radius of a circular orbit, and  ${\rm \tilde{L}}^{2} (r_0)>0$.
 
  Differentiation of identity $D{\rm \tilde{L}}^{2}+R^{3} \psi \left(A'\psi +2\psi '\right)\equiv 0$ in view of \eqref{L-of-r} at the point of minimum yields
\begin{equation} \label{GrindEQ__11_} 
 R^{3} U_{eff}''=-D\frac{d{\rm \tilde{L}}^{2} }{dr} ,\quad r=r_{0} ,    \end{equation} 
Thus, there are two conditions of SCO with radius $r$
\begin{equation}\label{SCO}
    {\rm \tilde{L}}^{2}(r) >0\quad {\rm and}\quad D(r)\frac{d{\rm \tilde{L}}^{2} }{dr}<0. 
 \end{equation}
 Obviously, the opposite signs correspond  either to UCO, or NECO. This makes it convenient in the numerical search for the   stability regions of circular orbits taking into account the slope of the graph ${\rm \tilde{L}}^{2} (r)$.

\section{Distributions of circular orbits}\label{sFJNW}
Introduction of the coupling of particles with SF creates additional contributions into the Einstein equations  (by means of the energy-momentum tensor)  and the SF equation (by means of the source which involves the particle masses). 
However, we avoid this question because we deal with the {test} particles moving in the external gravitational and scalar fields. This is a typical situation when  the contribution of the accretion disk is negligible compared to the mass of the supermassive object at the center of an  active galactic nucleus.

Under the requirements of the  spherical symmetry, there is the exact static F/JNW solution of the Einstein-SF equations  that describes the  massless linear SF in asymptotically flat space-time outside the naked singularity. We use this solution  in the form put by Virbhadra  \cite{Virbhadra_1997} 
\eq{A(r)=\left(1-\frac{r_s}{ r}\right)^{\gamma}=\frac{1}{B(r)},
\quad 
R^2(r)= r^2\left(1-\frac{r_s}{ r}\right)^{1-\gamma},\quad  \phi(r)=-\frac{Q}{r_s}\ln\left(1-\frac{r_s}{ r}\right)\label{f-JNW}, \quad r>r_s,}
where 
\begin{equation}
 r_s=2M/\gamma,\quad \gamma={\mu}/{\sqrt{\mu^2+1}},
\end{equation}
$M>0$ is the mass of the system,  $\mu=M/Q$,  $Q\in(-\infty, \infty), Q\ne 0$, is the scalar ``charge"\footnote{Our  notation differs from that of Chowdhury et al  \cite{Chowdhury_2012}, who define the scalar charge as $1/\gamma$.}, defined by the asymptotic dependence $\phi\sim Q/r$ for $r\to \infty$.   The radial coordinate $r$ is related to the coordinate $R$ of \cite{JNW} by $r=R+M(1+\gamma^{-1})$. For any $Q\ne 0$ there is NS for $r=r_s$; this is the center of curvature (Schwarzschild) coordinates. There is a critical value $\gamma_{c0}=1/2$ that separates two types of singularity at $r_s$; for $\gamma\in(\gamma_{c0},1)$ there exists the  root $r_{ph}=M(2+1/\gamma)$ of equation $D(r)=0$, which is the radius of the photon sphere; it does not depend on $\psi$.

Investigation  of \cite{Chowdhury_2012}  shows that (i) for $\gamma\in(0,\gamma_{c1})$, $\gamma_{c1}=1/\sqrt{5}$, there is  one domain of SCO with radii  $r>r_s$, which covers all the space outside NS; (ii)  for $\gamma\in(\gamma_{c1},\gamma_{c0})$ there are  two boundary radii  $r_1^{(b)}<r_2^{(b)}$ and two rings of SCO with $r\in (r_s,r_1^{(b)})$ and  $(r_2^{(b)},\infty)$,  and ring of UCO with  $r\in (r_1^{(b)},r_2^{(b)})$. (iii) For $\gamma\in(1/2,1)$ there is a ring of NECO with radii $r\in (r_s,r_{ph})$, then the ring of UCO with radii   $r\in (r_{ph},r_1^{(b)})$, and ring of SCO with  $r\in (r_1^{(b)},\infty)$.

The boundary radii of the SCO regions can be defined by the change of the sign either of $\tilde{L^2}(r^{(L)})$ or  of $d{\rm \tilde{L}}^{2}/dr$. Correspondingly,   throughout the paper we denote the roots of equations $\tilde{L^2}(r)=0$ by index ``$(L)$",   and the roots of $d{\rm \tilde{L}}^{2}/dr=0$ by index ``$(b)$" (on condition that $\tilde{L^2}(r^{(b)})>0$).

In subsections \ref{ssect:Power-law coupling} and \ref{exp-coupling} we  consider two simple examples of the interaction of particles with SF, which satisfy condition $\psi>0$, where $\psi$ contains an additional parameter $\kappa$ describing the strength of the coupling. Therefore, we have two parameters and the problem is to find areas of these parameters  in the $\gamma-\kappa$ plane and critical (bifurcation) curves that limit these areas. This is a typical problem of the catastrophe theory. There are two types of such curves. 

(a) The curves where the the regions of NECO with ${\rm \tilde{L}}^{2}(r)<0$ appear/disappear. They correspond to simultaneous fulfillment of equations (at the same $r$)
\begin{equation}\label{bifurk_L}
    {\rm \tilde{L}}^{2}(r,\gamma,\kappa)=0,\quad  d{\rm \tilde{L}}^{2}(r,\gamma,\kappa)/dr=0 .
\end{equation}

(b) The curves where the the minima/maxima of ${\rm \tilde{L}}^{2}(r)$ appear/disappear. They correspond to simultaneous fulfillment of equations  
\begin{equation}\label{bifurk_b}
    d{\rm \tilde{L}}^{2}(r,\gamma,\kappa)/dr=0,\quad  d^2{\rm \tilde{L}}^{2}(r,\gamma,\kappa)/dr^2=0 .
\end{equation}
However, in the numerical treatment it is more convenient to apply a straightforward determination of roots of ${\rm \tilde{L}}^{2}(r)$ and $d{\rm \tilde{L}}^{2}(r)/dr$. For admissible parameter values\footnote{In fact, in case of both couplings (\ref{PL-int}, \ref{Exp-int}) we tested the values of  $|\kappa|\in [0,10)$, but the region of $|\kappa|>1$  does not   result in qualitatively new elements and it is not shown in figures.}   we  look for possible  $r^{(L)}(\gamma,\kappa)$ and $r^{(b)}(\gamma,\kappa)$ and control the fulfillment of  conditions (\ref{SCO}) for SCO and analogous conditions for UCO and NECO for orbits with radii $r$ between these values.  
Namely, for  fixed $\gamma,\kappa$  we test the existence of  the solutions for $r>r_s$:  ${\rm \tilde{L}}^{2}(r)=0$ and/or $d{\rm \tilde{L}}^{2}/dr=0$, which separate $r$  corresponding to SCO, UCO and NECO.  The emergence and disappearance  of these solutions for some values of $\gamma,\kappa$  occurs on the bifurcation curves  separating the regions with different CODs in $\gamma-\kappa$ plane, colored in different colors in the Figures \ref{fig_1}, \ref{fig_2}, \ref{fig_3} below. The existence of different intervals for admissible radii of SCO and UCO is correspondingly shown in Tables \ref{table_1} and \ref{table_2} (see below).

\subsection{Power-law coupling}\label{ssect:Power-law coupling}
As mentioned above, we have no considerations other than  the simplicity criterion regarding the choice of $\psi(\phi)$ to formulate modifications of the relativistic gravity theory. The only restrictions must be imposed by  known weak-field experiments, which can be satisfied by an appropriate choice of model parameters that regulate the strength  of the interaction, not excluding qualitative  qualitative effects near the naked singularity. 
 In this situation, we will restrict ourselves to  the "toy" examples of the power-law coupling (this subsection) and exponential coupling (subsection \ref{exp-coupling}), which seem to be the simplest.

In the case of the power-law coupling 
 \eq{\label{PL-int}
 \psi(\phi)=1+\kappa \phi^{2n}\,\,,
 \quad n=1,2,...}
 where we consider the even powers of $\phi$ in (\ref{PL-int}) and we put $\kappa>0$ so as to have $\psi(\phi)>0$ for all $\phi$.
 
 From (\ref{L-of-r}) we have
 \begin{equation}\label{L-of-r_PL}
    {\rm \tilde{L}}^{2}(r)=\frac{r_s r  (1-r_s/r)^{1-\gamma}\psi F(r,Q,n)}{2(1-r_{ph}/r)}, \quad F(r)=\gamma+\kappa\left[\gamma \phi^{2n}-\frac{2n\gamma}{\mu}\phi^{2n-1}\right].
\end{equation}
 Without loss of the generality, we assume $Q>0$.
 
 We begin with some general analytical results concerning the sign of ${\rm \tilde{L}}^{2}(r)$ (i.e., where either SCO or UCO can exist) that follow directly from  (\ref{L-of-r_PL}).
 
 For a sufficiently large  $r$ we have   ${\rm \tilde{L}}^{2}(r)\approx rM $ and the conditions (\ref{SCO}) are fulfilled; therefore there is always  an unbounded outer region of SCO, which includes the non-relativistic orbits.

Now we proceed to the singularity neighborhood. 
For $r\to r_s$ we have $F(r)>0$. Therefore, ${\rm \tilde{L}}^{2}(r)>0$ is increasing at least in small vicinity of $r_s$ for $\gamma\in (0,1/2)$ (SCO) and there is NECO for $\gamma\in (1/2,1)$.

There is  bifurcation value $\kappa=\kappa_{cr}$, 
\begin{equation}\label{kappa_cr}
    \kappa_{cr}=\frac{\mu^{2n}}{(2n-1)^{2n-1}} = \frac{1}{(2n-1)^{2n-1}}\left( \frac{\gamma^2}{1-\gamma^2}\right)^{n}     ,
\end{equation}
 where the minimum of $F(r)$  changes its sign from positive for $\kappa\in(0,\kappa_{cr})$ to negative for  $\kappa>\kappa_{cr}$. This is the only minimum of $F(r)$, therefore for $\kappa>\kappa_{cr}$ we have only two zeros  of this function, which disappear at $\kappa_{cr}$. Equation (\ref{kappa_cr}) yields the bifurcation curve in the  $\gamma-\kappa$ plane corresponding to (\ref{bifurk_L}). 
 
 The following cases of behavior of ${\rm \tilde{L}}^{2}(r)$ are possible.

  (i) For $0<\kappa<\kappa_{cr}$ and $\gamma\in (0,1/2)$ we have ${\rm \tilde{L}}^{2}(r)>0$ for all $r$;
  
  (ii) For $0<\kappa<\kappa_{cr}$ and $\gamma\in (1/2,1)$ there are no circular orbits under the photon sphere and  ${\rm \tilde{L}}^{2}(r)>0$ for  $r>r_{ph}$;
 
 (iii) For $\kappa>\kappa_{cr}$  there are only two points $r_1^{(L)},\,r_2^{(L)}$, where   $F(r)$ changes its sign. Therefore, for $\gamma\in (0,1/2)$ we have  ${\rm \tilde{L}}^{2}(r)<0$ in $(r_1^{(L)},r_2^{(L)})$  (NECO) and ${\rm \tilde{L}}^{2}(r)>0$  otherwise. 
 
(iv) For $\kappa>\kappa_{cr}$ and $\gamma\in (1/2,1)$ the results depend upon mutual disposition of  $r_1^{(L)},\,r_2^{(L)}, r_{ph}$.

For $n=1$ the points  $r_1^{(L)},\,r_2^{(L)}$ can be easily found:
\begin{equation*}
   r_{1,2}^{(L)}=\frac{r_s}{1-e^{-\xi_{\pm}}} ,\quad \xi_{\pm}=\frac{2}{\gamma}\left(1\pm\sqrt{1-\frac{\mu^2}{\kappa}}\right), \quad \kappa> {\mu^2} .
\end{equation*}

\begin{figure}
    \centering
\includegraphics[width=.49\textwidth]{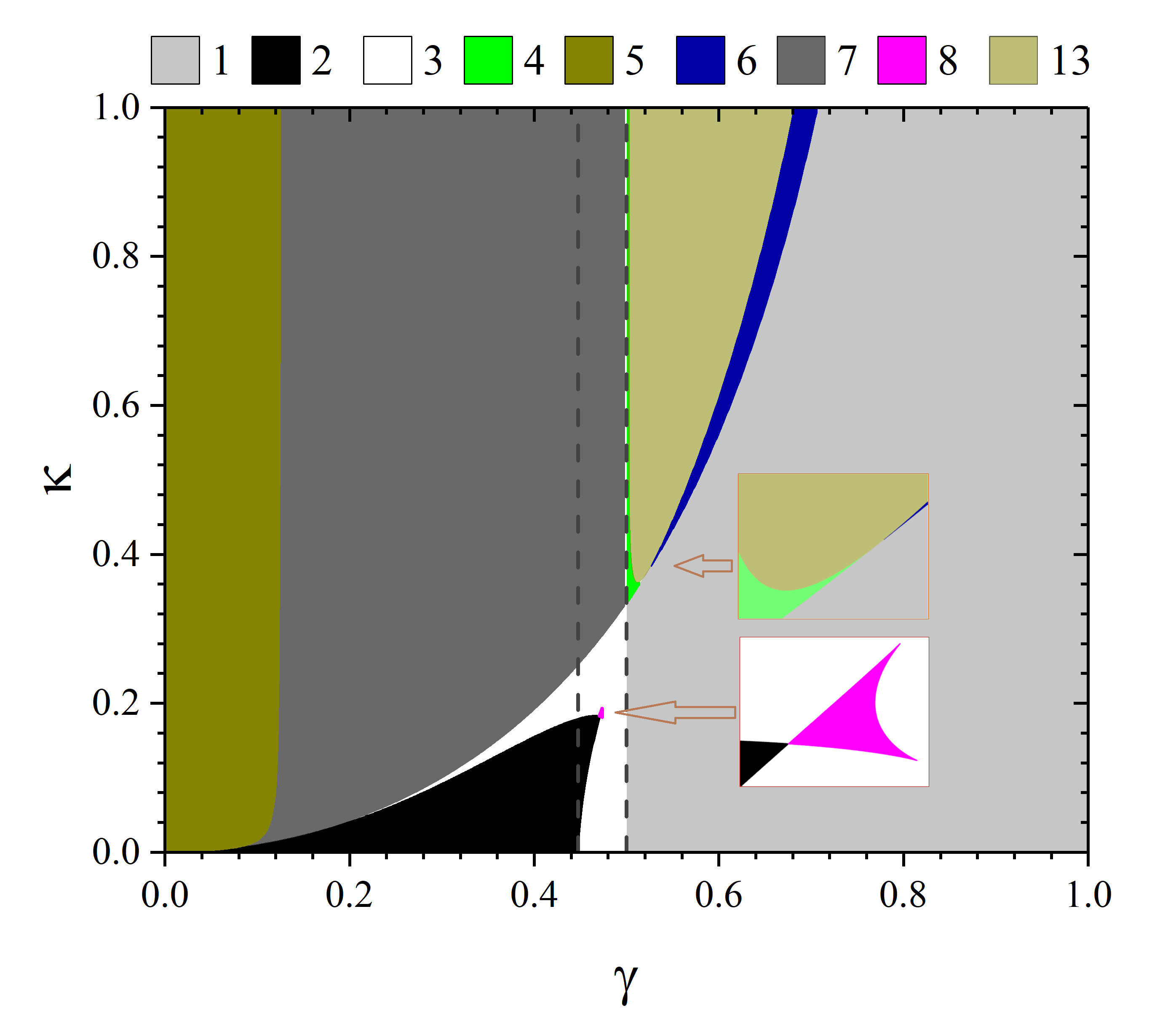}
\includegraphics[width=.49\textwidth]{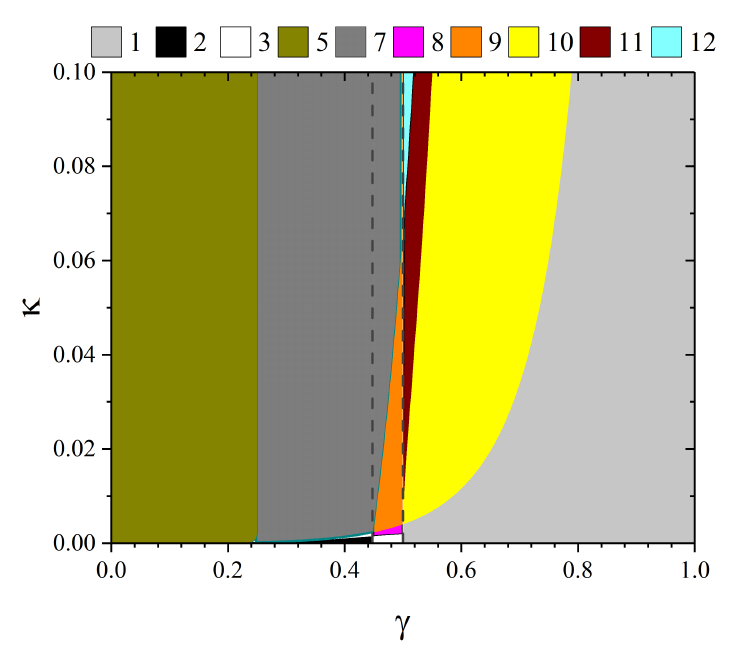}
\caption{Possible COD cases for the coupling  (\ref{PL-int}). Left panel: $n=1$, right panel $n=2$.  The dashed vertical lines correspond to critical values of  $\gamma_{c0}=1/2$ and $\gamma_{c1}=1/\sqrt{5}$. The rectangles in the left panel represent enlarged elements, shown by the arrows.}
    \label{fig_1}
\end{figure}

We studied numerically function ${\rm \tilde{L}}^{2}(r)$ for  $n=1,2$. Fig. \ref{fig_1} shows areas with qualitatively different  COD  in $\gamma-\kappa$ plane which correspond Table \ref{table_1}. Here we confined ourselves with $0\le \kappa\le 1$, which is sufficient to see the general tendency of the results.  The areas of Fig. \ref{fig_1} with qualitatively  different SOD are coloured in different colours indicated in the legend of each of two panels with the number associated with  certain row of the Table  \ref{table_1}. Every row shows  intervals of the radii of possible SCO and UCO regions,   in accordance with the indexing introduced   above). 

\begin{figure}
    \centering
\includegraphics[width=.49\textwidth]{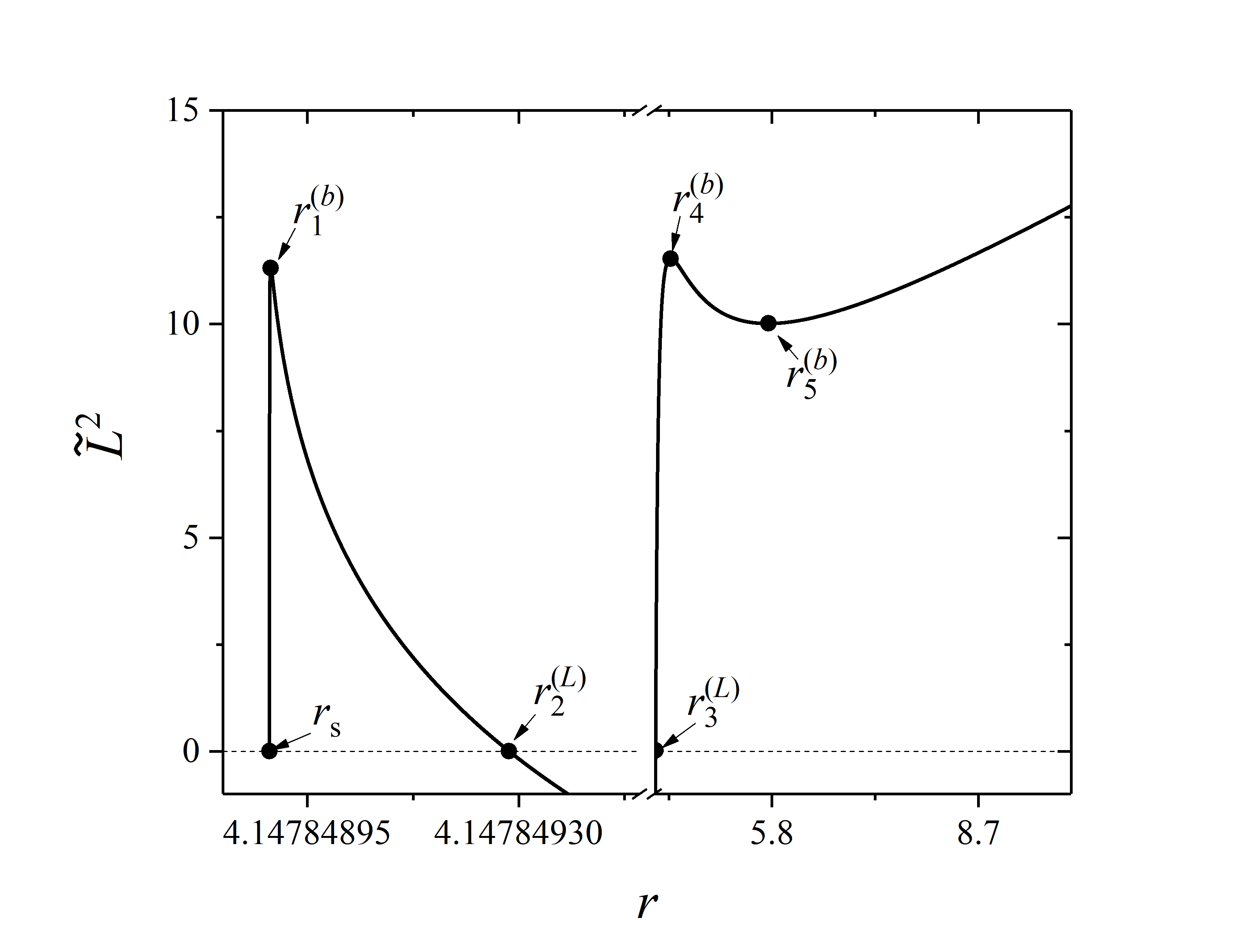}
\includegraphics[width=.49\textwidth]{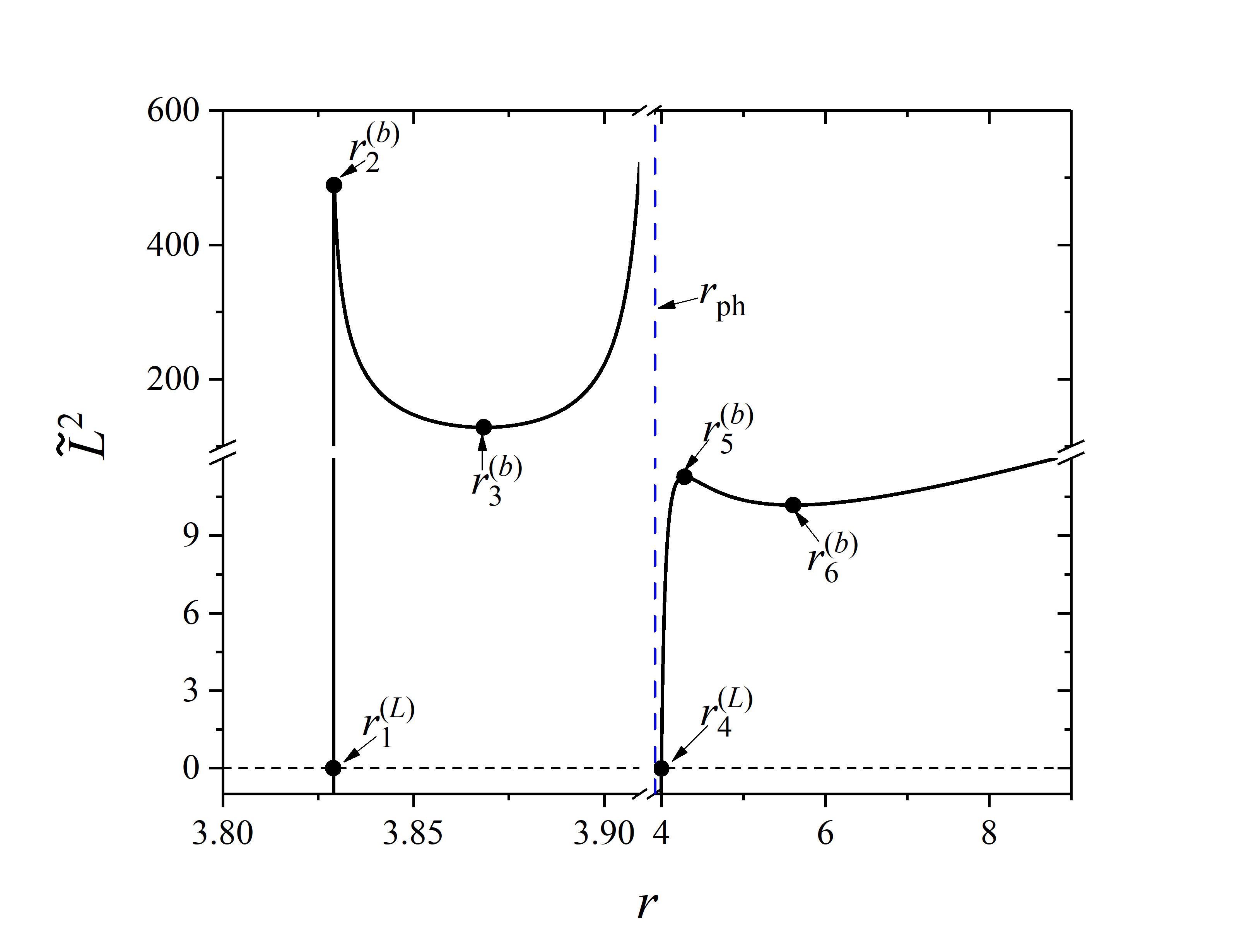}
\caption{Typical behaviour of ${\tilde L}^2$ for area 9 (left)   and area 11  (right) of the Table \ref{table_1} and Fig. \ref{fig_1} in case of monomial coupling (\ref{PL-int}). There is very fast growth of ${\tilde L}^2$ and very small size of the SCO regions near the singularity. For case 9: $r_1^{(b)}-r_s\simeq2\cdot10^{-9}$; for case 11: $r_2^{(b)}-r_1^{(L)}\simeq0.0001$.}
    \label{figure_graph}
\end{figure}

As an example, we explain the notations  in case of the orange area (number 9 of the legend) in the right panel ($n=2$) of Fig. \ref{fig_1}. Correspondingly, row 9 of  Table \ref{table_1} shows that there exist  boundary radii $r_1^{(b)},r_2^{(L)},r_3^{(L)},r_4^{(b)}, r_5^{(b)}$ of regions with different stability properties (in ascending order of the lower indexes), which of course depend on $\gamma$ and $\kappa$ inside the orange area. Here the coordinate $r_s$ of the singularity is also involved. The  left column (row 9) shows  that  for every  $\gamma$ and $\kappa$ of this area, there exist such values of  boundary radii  that function ${\rm \tilde{L}}^{2}(r)>0$ is increasing  in three intervals  $(r_s,r_1^{(b)})$, $(r_3^{(L)},r_4^{(b)})$, $(r_5^{(b)},\infty)$, where we have SCO (see Fig. \ref{figure_graph}, left panel). This function decreases in $(r_1^{(b)},r_2^{(L)})$ and $(r_4^{(b)},r_5^{(b)})$ (see row 9, middle column of  Table \ref{table_1}, UCO). In the remaining intervals, not explicitly shown in Table \ref{table_1}, we have NECO. The right column  indicates that $\gamma<1/2$, i.e. the orange region is to the left of the vertical dashed line in the right panel of Fig. \ref{fig_1}.   

Fig. \ref{figure_graph} (right panel)  shows the other example of the graph of ${\rm \tilde{L}}^{2}(r)$, which is related to row 11 of Table \ref{table_1} (brown region   in the right panel of Fig. \ref{fig_1},  to the right of the dashed line  $\gamma=1/2$).

\begin{table}[htbp] 
\begin{tabular}{|c|c|c|c|c|}
\hline
N &SCO  &UCO    & $sign(\gamma- 0.5)$  \\ \hline
1 &$(r_1^{(b)},\infty) $ & $(r_{ph},r_1^{(b)})$  & $+$  \\ \hline
2 &$(r_s,\infty) $ & $-$  & $-$  \\ \hline
3 &$(r_{s},r_{1}^{(b)})\cup(r_2^{(b)},\infty)$  &$(r_{1}^{(b)},r_2^{(b)})$  & $-$ \\ \hline
4&$(r_2^{(L)},\infty)$  & $(r_{ph},r_1^{(L)})$& $+$\\ \hline
5&$(r_1^{(L)},\infty)$  & $-$& $+/-$\\ \hline
6&$(r_2^{(b)},r_3^{(L)})\cup(r_4^{(b)},\infty)$  & $(r_1^{(L)},r_2^{(b)})\cup(r_{ph},r_4^{(b)})$& $+$\\ \hline
7&$(r_s,r_1^{(b)})\cup(r_3^{(L)},\infty)$  & $(r_1^{(b)},r_2^{(L)})$& $-$\\ \hline
8&$(r_s,r_1^{(b)})\cup(r_2^{(b)},r_3^{(b)})\cup(r_4^{(b)},\infty)$  & $(r_1^{(b)},r_2^{(b)})\cup(r_{3}^{(b)},r_4^{(b)})$& $-$\\ \hline
9&$(r_s,r_1^{(b)})\cup(r_3^{(L)},r_4^{(b)})\cup(r_5^{(b)},\infty)$  & $(r_1^{(b)},r_2^{(L)})\cup(r_{4}^{(b)},r_5^{(b)})$& $-$\\ \hline
10&$(r_2^{(b)},r_3^{(L)})\cup(r_4^{(b)},\infty)$  & $(r_1^{(L)},r_2^{(b)})\cup(r_{ph},r_4^{(b)})$& $+$\\ \hline
11&$(r_2^{(b)},r_3^{(b)})\cup(r_4^{(L)},r_5^{(b)})\cup(r_6^{(b)},\infty)$  & $(r_1^{(L)},r_2^{(b)})\cup(r_3^{(b)},r_{ph})\cup(r_{5}^{(b)},r_6^{(b)})$& $+$\\ \hline
12&$(r_3^{(b)},r_4^{(b)})\cup(r_4^{(L)},\infty)$  & $(r_1^{(L)},r_2^{(b)})\cup(r_3^{(b)},r_{ph})$& $+$\\ \hline
13&$(r_2^{(L)},\infty)$  & $(r_1^{(L)},r_{ph})$& $+$\\ \hline
\end{tabular}
\caption{Possible SCO and UCO regions for monomial coupling (\ref{PL-int}) in correspondence  to different areas of Fig. \ref{fig_1}.}
\label{table_1}
\end{table}

The case  $\gamma=1/2$ is poorly visible in the figures and we describe it separately.  In this case   $D(r)<0$ for all $r>r_s$. 

For $n=1$, there is a critical value $\kappa_{cr}\approx 0.33$  such that for $\kappa<\kappa_{cr}$   $\exists r_1^{(b)}$ such that ${\rm \tilde{L}}^{2}(r)>0$ decreases for $r\in(r_{ph},r_1^{(b)})$ corresponding to UCO region near the center,  and increases  for $r>r_1^{(b)}$ (SCO). Here  $r_1^{(b)}$ is the point of minimum of ${\rm \tilde{L}}^{2}$; this minimum decreases as $\kappa$ grows and becomes negative after   $\kappa>\kappa_{cr}$, when $\exists r_1^{(L)}, r_2^{(L)}$ such that there is UCO region with radii $r\in(r_{ph},r_1^{(L)})$, NECO for $r\in( r_1^{(L)},r_2^{(L)})$ and SCO for $r>r_2^{(L)}$.

In the case  $\gamma=1/2$, $n=2$, we have  analogous situation; the critical $\kappa_{cr}$ can be found from (\ref{kappa_cr}). 

 \subsection{ Exponential coupling}\label{exp-coupling}
Here we consider an exponential type of  the coupling, inspired by considerations from \cite{Damour-1993, Shtanov:2021} in the form of 
\eq{
 \label{Exp-int}\psi(\phi)=e^{\kappa \phi^{n}};\quad  \xi=e^{\kappa \phi^{n}}-1,~n= 1,2,..., 
 }
now $\psi(\phi)>0$ for all values of $\phi$  and $\kappa\in(-\infty,\infty)$. Note that unlike subsection \ref{ssect:Power-law coupling},  the sign of $Q$ is significant for odd $n$, however, in this case instead of considering negative $Q$ one can change $\kappa\to -\kappa$; therefore, we always assume $Q>0$ and correspondingly $\phi>0$.

We have from (\ref{L-of-r}) 
\begin{equation}
   {\rm \tilde{L}}^{2} =rM  \frac{(1-r_s/r)^{1-\gamma}}{1-r_{ph}/r} \left(1-  \frac{\kappa n }{\mu}  \phi^{n-1}  \right)e^{2\kappa\phi^n}.
\end{equation}
For $\kappa<0$ we always have ${\rm \tilde{L}}^{2}>0$ for $r>r_{ph}$, if $\gamma\in (1/2,1)$ and NECO for $r_s<r<r_{ph}$; and ${\rm \tilde{L}}^{2}>0,\,\forall r>r_s$ if  $\gamma\in (0,1/2)$.  

For $n>1$ and $\kappa>0$, the root of equation ${\rm \tilde{L}}^{2}=0$ can be written as
\[
r^{(L)}=\frac{1}{1-e^{-\xi}},\quad
\xi=2\left(\frac{\gamma}{\kappa n }\right)^{1/(n-1)}\left[\frac{1}{\sqrt{1-\gamma^2}}\right]^{n/(n-1)}>0. 
\]
These roots are always uniquely defined, no appearance/disappearance of roots does not occur. Therefore, there are no bifurcation curves of type (a) (see (\ref{bifurk_L})).

\begin{figure}
    \centering
\includegraphics[width=.49\textwidth]{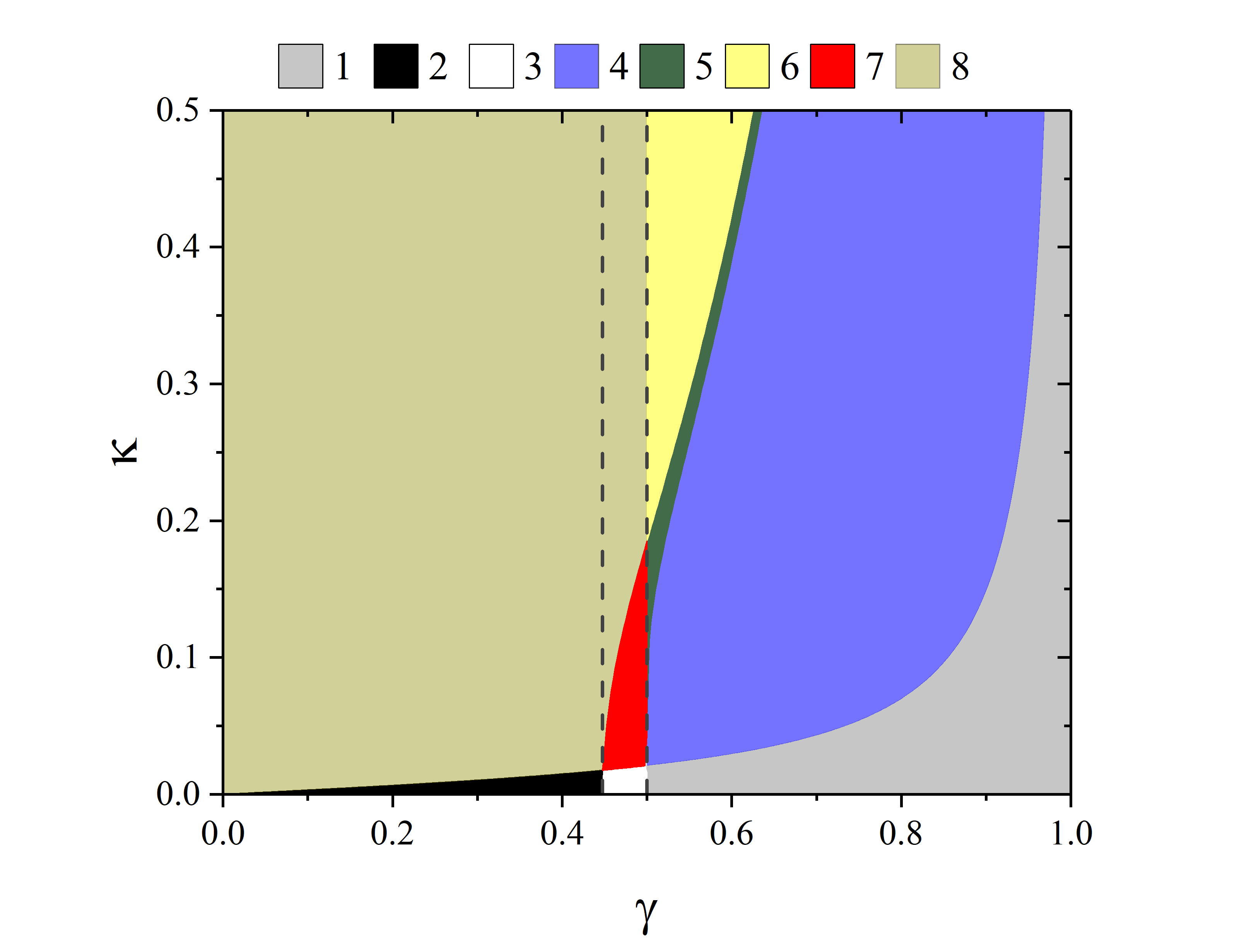}
\includegraphics[width=.49\textwidth]{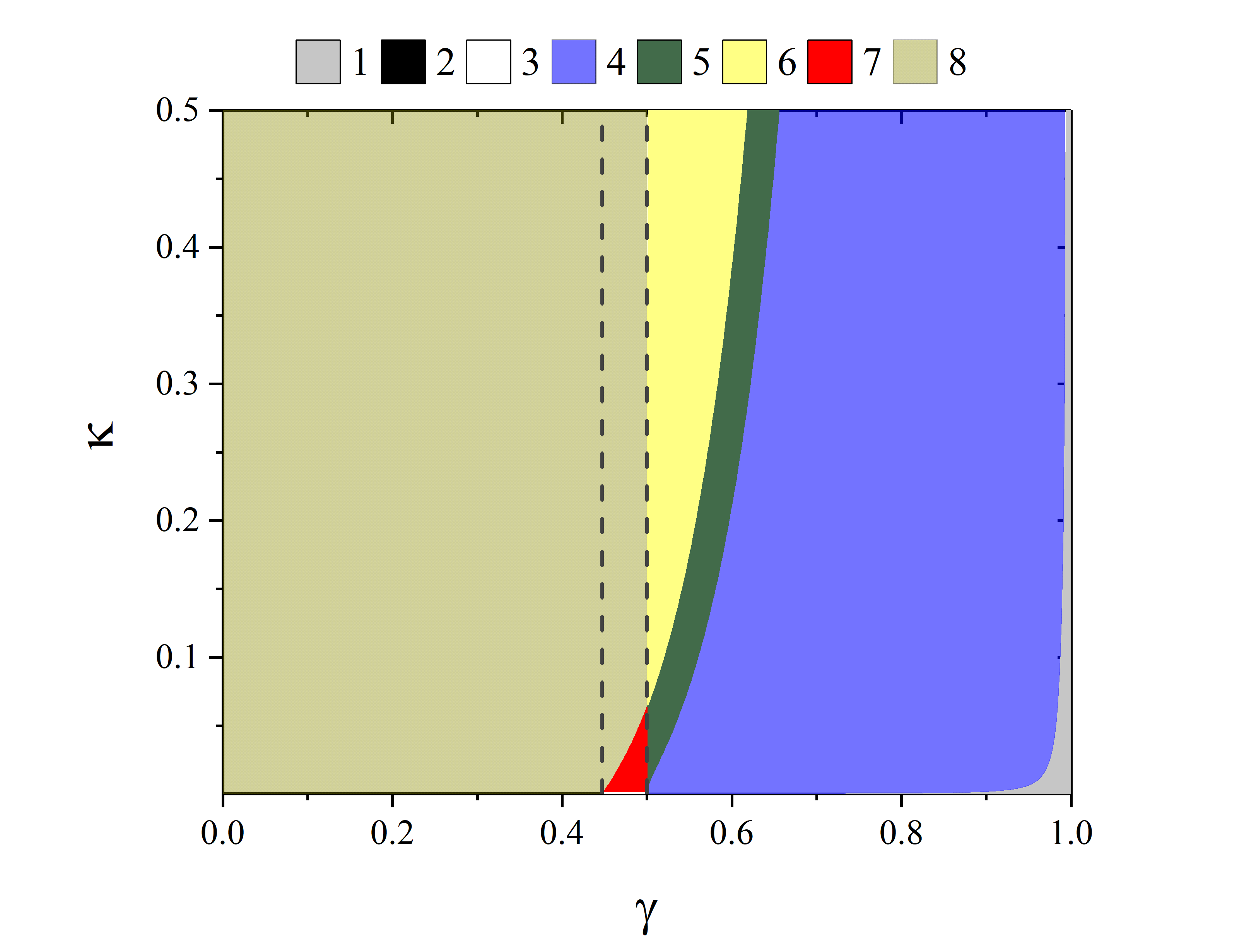}
\caption{Possible COD cases for (\ref{Exp-int}), $\kappa>0$.   Left panel: $n=2$, right panel $n=4$.  The dashed vertical lines as in Fig.\ref{fig_1}.}
    \label{fig_2}
\end{figure}

\begin{figure}
    \centering
\includegraphics[width=.49\textwidth]{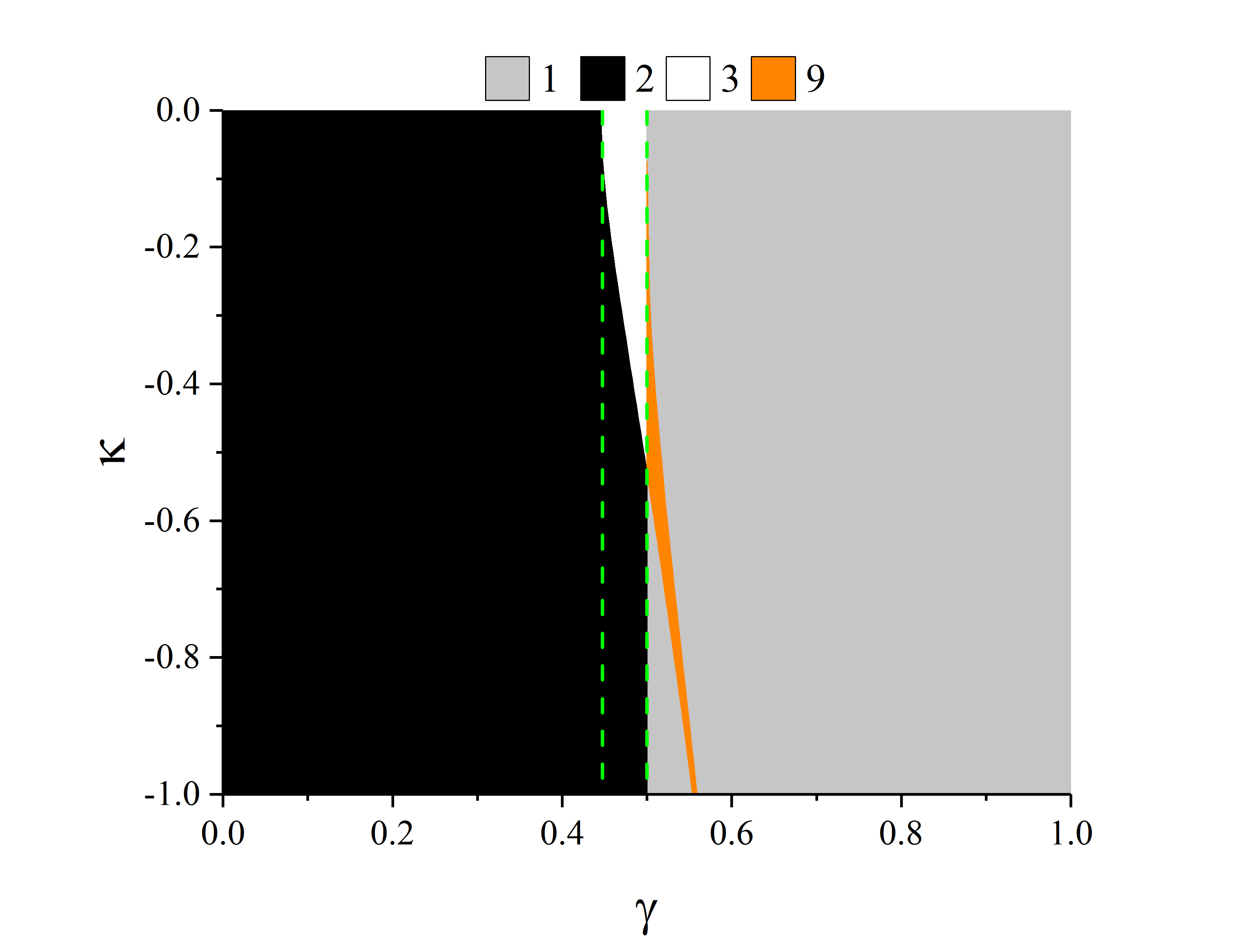}
\includegraphics[width=.49\textwidth]{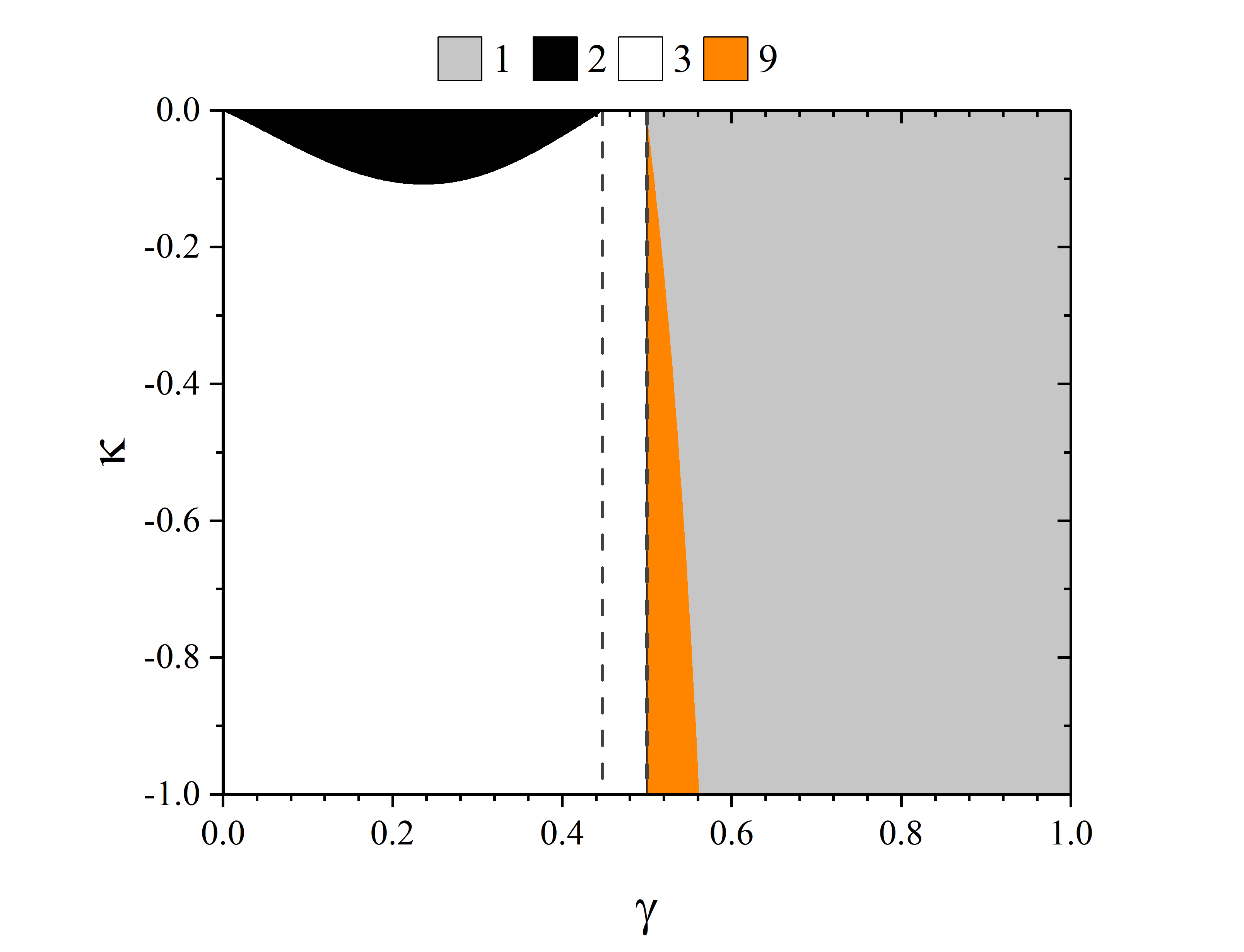}
\caption{Possible COD cases for (\ref{Exp-int}), $\kappa<0$.   Left panel: $n=2$, right panel $n=4$.  The dashed vertical lines as in Fig.\ref{fig_1}.}
    \label{fig_3}
\end{figure}

Further information on COD obtained numerically is represented in Figs. \ref{fig_2}, \ref{fig_3} and Table  \ref{table_2}. The latter contains one more column indicating the possible sign of $\kappa$. 

In case of $n=1$ a detailed analytic treatment is possible. There is an obvious bifurcation curve
\[
\kappa_{b,0}=\mu=\gamma/\sqrt{1-\gamma^2}, 
\]
that separates regions in the $\gamma-\kappa$ plane:
\begin{itemize}
\item  For $\kappa <\mu$, $\gamma\in (0,1/2)$   we always have ${\rm \tilde{L}}^{2}>0,\,\forall r>r_s$.
\item  For   $\gamma\in (1/2,1)$   we  have ${\rm \tilde{L}}^{2}>0$ (UCO or SCO) for  $r>r_{ph}$ and NECO for $r_s<r<r_{ph}$. 
\item There is a reverse situation for $\kappa >\mu$. In particular, there is no circular orbits at all for $\gamma\in (0,1/2)$; in this case the repulsion of particles from the center due to SF overcomes usual gravitational attraction.  For $\gamma\in (1/2,1)$ we have ${\rm \tilde{L}}^{2}>0$ for $r<r_{ph}$. There is SCO for $r\in(r_s,r_1^{(b)})$, UCO for $r\in(r_1^{(b)},r_{ph})$ and NECO for $r>r_{ph}$. The situations with NECO for large $r$ are obviously  unphysical.

\item
From the conditions (\ref{extremum_U},\ref{GrindEQ__11_}) we obtain analytic expression for two boundary radii 
\eq{r^{(b)}_{1,2}=\gamma^{-1} \left[{1+3 \gamma +\sqrt{1-\gamma ^2} \kappa\pm\sqrt{\gamma ^2 \left(5- \kappa ^2 \right)+2
   \sqrt{1-\gamma ^2} \gamma  \kappa +\kappa
   ^2-1} }\right] \,\,.}
   These roots disappear on the type (b)  bifurcation curve   (see (\ref{bifurk_b})) with branches
 \begin{equation}
    \kappa_{1,2}=\frac{\gamma\pm\sqrt{1-4\gamma^2}}{\sqrt{1-\gamma^2}}= \mu\pm 1 \,,  \quad \gamma<\frac12 \,.
 \end{equation}
 If we include into consideration the negative $\gamma$ and $\kappa$, these branches form the closed curve in the $\gamma-\kappa$ plane, which cross the $\gamma$--axis at $\gamma=\pm 1/\sqrt{5}$.
\end{itemize}

\begin{table}[htbp]
\begin{tabular}{|c|c|c|c|c|}
\hline
N &SCO  &UCO  &$sign(\kappa)$  &$sign(\gamma- 0.5)$  \\ \hline
1 &$(r_{1}^{(b)},\infty)$&$(r_{ph},r_1^{(b)})$   & $+/-$ & $+$ \\
\hline
2 &$(r_{s},\infty)$ & $-$ &$+/-$ & $-$ \\ \hline
3 &$(r_{s},r_{1}^{(b)})\cup(r_2^{(b)},\infty)$&$(r_{2}^{(b)},r_3^{(b)})$ & $+/-$ & $-$ \\ \hline
4 &$(r_{s},r_{1}^{(L)})\cup(r_{2}^{(b)},\infty)$&$(r_{ph},r_2^{(b)})$   & $+$ & $+$ \\ \hline
5 &$(r_s,r_1^{(b)})\cup(r_2^{(L)},r_3^{(b)})\cup(r_4^{(b)},\infty) $  &$(r_1^{(b)},r_{ph})\cup(r_3^{(b)},r_4^{(b)})$  & $+$ & $+$ \\ \hline
6 &$(r_s,r_1^{(b)})\cup(r_2^{(L)},\infty)$&$(r_{1}^{(b)},r_{ph})$   & $+$ & $+$ \\
\hline
7 &$(r_1^{(L)},r_2^{(b)})\cup(r_3^{(b)},\infty) $ & $(r_2^{(b)},r_3^{(b)})$ & $+$ & $-$  \\ \hline
8 &$(r_1^{(L)},\infty) $ & $-$  & $+$ & $-$  \\
\hline
9 &$(r_{1}^{(b)},r_{2}^{(b)})\cup(r_3^{(b)},\infty)$& $(r_{ph},r_1^{(b)})\cup(r_2^{(b)},r_3^{(b)})$ & $-$ & $-$\\ \hline
\end{tabular}
\caption{Possible SCO and UCO regions for the exponential coupling (\ref{Exp-int}) in correspondence  with  different areas of Figs. \ref{fig_2}, \ref{fig_3}.}
\label{table_2}
\end{table}

\section{Discussion}\label{discussion}
We have shown that introduction of the coupling of particles with SF can essentially change the SCO distribution around the naked singularity described by the F/JNW solution and creates a wealth of new types. In particular, COD with three SCO regions  appear in case of the non-trivial coupling, which are absent in case of F/JNW.  Namely, in case the geodesic motion ($\kappa=0$), when the test particles move along geodesics in the F/JNW metric, we have only one case of disjoint SCO regions. In case  of the monomial coupling (\ref{PL-int}), we found  areas in the $\gamma-\kappa$ plane corresponding  to three SCO regions (see lines 8, 9 and 11) of Table \ref{table_1}), which are separated by UCO/NECO rings. In case of the exponential coupling analogous case  corresponds to line 5 of Table \ref{table_2}.

Let us briefly discuss possible applications of  the obtained results to astrophysical objects.

The relation of the SCO distributions to the accretion disks around compact objects is straightforward  in the Page - Thorn model \cite{Page_Thorne}. Moreover, one can expect that main  signatures of the  SCO rings,  separated by the instability regions, will retain their relevance in more complex models of accretion disks as well. 
For example, the  dark spot at the center of an accretion disk around a black hole is ultimately associated with the NECO and UCO regions, although the underlying physical processes in the disk can be very complex. Thus, if the configurations with SF really exist, then the  disconnected structure of the SCO distributions may be an important  feature that can be related to the observational properties of the accretion discs in astrophysical systems.

However, the occurrence of the most interesting COD structures  require a fine tuning, because they occupy rather small areas in the $\gamma-\kappa$ plane. Also, it is evident that the existence of such fine details is possible in the region of strong gravitational field that is not far from the singularity, where the "unusual" annular structures are typically very small and they might be  difficult to observe.

Is there  restrictions of the weak-field gravitational experiments in connection with the above results? 
The Post-Newtonian approximation of the metric (\ref{f-JNW}) in the isotropic coordinates does not involve the scalar charge $Q$, so the  PPN-parameters of F/JNW solution are the same as in the Schwarzschild case \cite{Formida}. Therefore, if there is no interaction of particles with SF, then in the case when the values of  $M$  and $Q$ are of  the same order, the latter practically is not restricted by the Solar system gravitational  experiments. 

The situation changes, if the interaction is switched on. The post-Newtonian analysis can be easily carried out thanks to the fact that the equations  (\ref{Eqom})  can be considered  as the geodesics in the space-time with the conformal metric  $\tilde g_{\mu\nu}=\psi^2 g_{\mu\nu}$ ($\psi>0$). In the cases described by (\ref{PL-int}) and (\ref{Exp-int}), for $|\xi|<<1$, we have $\xi\sim \kappa |\phi|^n\sim \kappa (|Q|/r)^n$. 

 For both models (\ref{PL-int}) and (\ref{Exp-int}), the constant $\kappa$ should be considered universal. Then, if  $n=1$ in the case of (\ref{PL-int}) or $n=1,2$ in the case of (\ref{Exp-int}),  for $\kappa\sim 1$  the Solar system  experiments put rather strong restrictions on the scalar charge of  the  system.  Otherwise, one must consider $\kappa<< 1$ and  the properties of SCO distributions will be essentially the same as for $\kappa=0$, even if $Q$ is considerable. 
There seems to be nothing unusual about  the images of accretion disks around M87* and SgrA* provided by the Event Horizon Telescope (EHT) team \cite{EHT-2019, EHT-2022}. Therefore, it may seem that the case of significant $Q$ in these systems with $\kappa\sim 1$ can be excluded. However, it should be kept in mind that  the current resolution of  EHT   may be  insufficient to resolve the discussed above  (putative) fine structure of the accretion discs (cf. \cite{Miyoshi2022}). Also, the absence of such a structures in M87* and SgrA* does not mean that they cannot be present in other objects.

For larger $n$ and  $\kappa \sim 1$, the scalar charge does not to contribute into the Post-Newtonian orders and therefore, to the PPN-parameters; then the Solar system experiments do not restrict $Q$ and $\kappa$.
 In this case we have a sufficient freedom to chose $Q$ and $\kappa$.

Although  F/JNW is a very special solution dealing with massless linear SF, one can suppose that situation with several non-connected SCO regions represents more general case. For example, it was pointed out in \cite{SZA2}  that the circular orbits for static spherically symmetric configurations with the SF potential $\sim \phi^{2p}$, $p>2$, have much in common with the F/JNW case. We expect that  the appearance  of new COD types after switching on the coupling of  particles with SF occurs in this case as well. 
\\

\textbf{Acknowledgements}
We are grateful to the anonymous referee for helpful comments, which stimulated the improvement of the manuscript. O.S. is thankful to Horst Stöcker for the warm hospitality at Frankfurt Institute for Advanced Studies.
This work has been supported in part by the
scientific program “Astronomy and space physics” of
Taras Shevchenko National University of Kyiv, project 22BF023-01.

\bibliography{references.bib}
 
\end{document}